\documentclass[9pt,twocolumn,twoside]{opticajnl}

\usepackage{lineno}

\usepackage{tikz}
\usepackage{bm}
\usepackage{hyperref}
\usepackage{multirow}
\usepackage{pifont}
\usepackage{mathtools}
\usepackage{pgfplots}
\usepackage{amsmath}

\usepgfplotslibrary{groupplots,dateplot}
\pgfplotsset{compat=newest}
\usetikzlibrary{patterns,shapes,arrows,arrows.meta,chains,fit,positioning,calc}

\DeclareMathOperator*{\argmin}{\arg\!\min}

\newcommand{\cmark}{\ding{51}}%
\newcommand{\xmark}{\ding{55}}%

\graphicspath{{../Figures/}}
\colorlet{lcfree}{black}
\colorlet{lcnorm}{black}
\colorlet{lccong}{black}

\newcommand{\fig}{Fig.\ }
\newcommand{\T}{^\text{T}}

\newcommand{\Height}{H}
\newcommand{\Width}{W}
\newcommand{\xsrc}{x^{\text{src}}}
\newcommand{\ysrc}{y^{\text{src}}}
\newcommand{\xdst}{x^{\text{dst}}}
\newcommand{\ydst}{y^{\text{dst}}}
\newcommand{\xisrc}{x_i^{\text{src}}}
\newcommand{\yisrc}{y_i^{\text{src}}}
\newcommand{\xidst}{x_i^{\text{dst}}}
\newcommand{\yidst}{y_i^{\text{dst}}}
\newcommand{\Homography}{\bm{H}}

\newcommand{\disparity}{d}
\newcommand{\baseline}{b}
\newcommand{\focallength}{f}
\newcommand{\depth}{p}
\newcommand{\pixelsize}{s}

\newcommand{\LightSource}{q}
\newcommand{\Reflectance}{r}
\newcommand{\CameraLens}{o}
\newcommand{\CameraSensor}{m}
\newcommand{\Wavelength}{\lambda}
\newcommand{\HSImageSpectrum}{s}

\newcommand{\Block}{\bm{B}}
\newcommand{\x}{\bm{x}}
\newcommand{\y}{\bm{y}}
\newcommand{\ReferenceVector}{\bm{r}}
\newcommand{\DistortedVector}{\bm{d}}
\newcommand{\LinRegA}{\alpha}
\newcommand{\LinRegB}{\beta}
\newcommand{\BestMatches}{B}

\newcommand{\FilterMatrix}{\bm{F}}
\newcommand{\MSPixel}{\bm{c}}
\newcommand{\Filter}{\bm{f}}
\newcommand{\HSPixel}{\bm{s}}

\newcommand{\PRIBPBlockPixel}{b}
\newcommand{\PRIBPRefPixel}{r}
\newcommand{\PRIBPRef}{\bm{r}}
\newcommand{\PRIBPBand}{\bm{b}}
\newcommand{\PRIBPKnownRef}{\tilde{\bm{r}}}
\newcommand{\PRIBPKnownBand}{\tilde{\bm{b}}}

\newcommand{\InterWidth}{M}
\newcommand{\InterHeight}{N}
\newcommand{\InterChannel}{C}
\newcommand{\InterMotion}{\bm{V}}
\newcommand{\InterPrediction}{P}
\newcommand{\InterResidual}{R}
\newcommand{\InterMask}{M}
\newcommand{\InterCoder}{\text{C}}
\newcommand{\InterDecoder}{\text{D}}

\fancypagestyle{firststyle}
{
	\fancyhf{}
	\fancyfoot[L]{\copyright 2023 Optica Publishing Group. One print or electronic copy may be made for personal use only. Systematic reproduction and distribution, duplication of any material in this paper for a fee or for commercial purposes, or modifications of the content of this paper are prohibited.}
}

\journal{josaa}

\setboolean{shortarticle}{false}

\title{Synthetic Hyperspectral Array Video Database with Applications to Cross-Spectral Reconstruction and Hyperspectral Video Coding}

\author[1, *]{Frank Sippel}%
\author[1]{Jürgen Seiler}%
\author[1]{André Kaup}%

\affil[1]{Friedrich-Alexander-Universität Erlangen-Nürnberg (FAU), Cauerstr. 7, 91058 Erlangen, Germany}

\affil[*]{Corresponding author: frank.sippel@fau.de}

\doi{\url{https://doi.org/10.1364/JOSAA.479552}}

\begin{abstract}
In this paper, a synthetic hyperspectral video database is introduced.
Since it is impossible to record ground truth hyperspectral videos, this database offers the possibility to leverage the evaluation of algorithms in diverse applications.
For all scenes, depth maps are provided as well to yield the position of a pixel in all spatial dimensions as well as the reflectance in spectral dimension.
Two novel algorithms for two different applications are proposed to prove the diversity of applications that can be addressed by this novel database.
First, a cross-spectral image reconstruction algorithm is extended to exploit the temporal correlation between two consecutive frames.
The evaluation using this hyperspectral database shows an increase in PSNR of up to 5.6 dB dependent on the scene.
Second, a hyperspectral video coder is introduced which extends an existing hyperspectral image coder by exploiting temporal correlation.
The evaluation shows rate savings of up to 10\% depending on the scene.
The novel hyperspectral video database and source code is available at \textit{\url{https://github.com/FAU-LMS/HyViD}} for use by the research community.
\end{abstract}

\setboolean{displaycopyright}{true}

\begin{document}

\maketitle
\thispagestyle{firststyle}

\section{Introduction}
\label{sec:introduction}
Classical cameras record a scene using three color channels, namely, the red (R), green (G) and blue (B) channel.
To achieve this, the incoming light spectrum is filtered by corresponding filters, e.g., a filter for the red component lets pass the wavelengths from 600 nm to 700 nm.
The RGB filters are approximating the properties of the different cones of the human eye.
Thus, if such an image is shown to a human on a proper display, all perceivable information is recorded and displayed.

For many classification tasks, it is useful to record even more channels, for example also in the infrared (IR) area of the spectrum.
Thus, the concept of RGB imaging is extended to \textit{multispectral} imaging, where typically six to 16 different bands of the scene are recorded.
Usually, the filters of multispectral cameras are spectrally distributed depending on the application.
Therefore, there is no structure regarding the shape of filters or the location of filters in the spectrum.

In contrast, the filters of \textit{hyperspectral} cameras are typically equidistantly distributed in the wavelength area of interest.
Since the employed bandpass filters are narrowband as well, the goal of a hyperspectral camera is to sample the light spectrum in the desired wavelength area for each pixel.
While there are no sharp definitions of multispectral and hyperspectral, these definitions are inline with the definitions given by Hagen et al.\cite{hagen_review_2013}, which state that most authors distinguish between contiguous (hyperspectral) versus spaced spectral bands (multispectral).

Hyperspectral imaging has become increasingly popular for diverse classification tasks.
For example, different types of plastics can be separated using hyperspectral cameras\cite{garaba_airborne_2018}, which is an essential building block for recycling pipelines.
Moreover, these cameras are employed in agriculture to measure the plant health\cite{williams_classification_2016} in order to distribute the proper amount of water and fertilizer.
Other applications include forensics\cite{edelman_hyperspectral_2012} where hyperspectral cameras can analyse the scene in a non-invasive way, object tracking\cite{xiong_material_2020}, and tumour detection in medicine\cite{han_vivo_2016}.
Usually, for most of these applications, it is beneficial if the hyperspectral camera is able to record videos.

\begin{figure}
	\centering
	\includegraphics[]{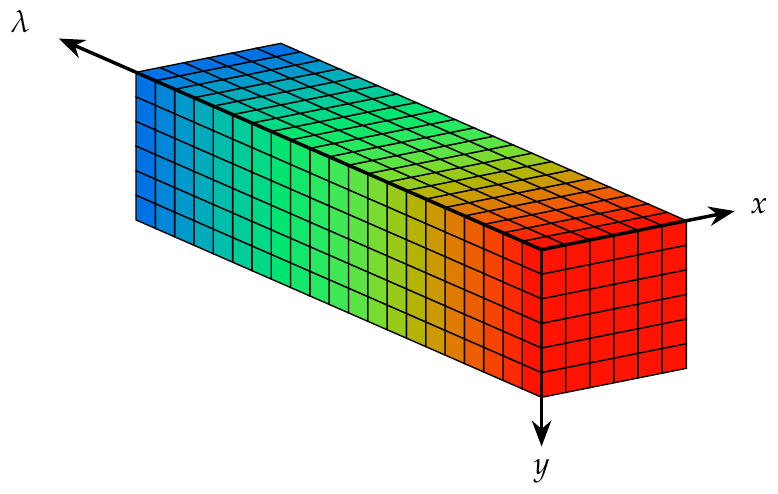}
	
	\caption{A hyperspectral 3D datacube with two spatial dimensions $x$ and $y$ and one spectral dimension $\lambda$.}
	\label{fig:hs_cube}
\end{figure}

\begin{table*}[t]
	\centering
	\renewcommand{\arraystretch}{1.5}
	\caption{Overview over different hyperspectral databases. Note that the 1890 images of the proposed database are put together by 7 scenes, each with 30 frames, rendered from a camera array with 9 cameras.}
	\label{tab:databases}
	\begin{tabular}{ccccccccc}
		Database                                        & Spatial res. & \#Channels & Wavel. area (nm) & \#Images & Videos & Rerendering & Cam. array & Depth   \\ \hline
		CAVE\cite{yasuma_generalized_2010}              & $512 \times 512$      & 31    & 400 - 700     & 32   & \xmark    & \xmark    & \xmark & \xmark \\
		UGR\cite{eckhard_outdoor_2015}                  & $1000 \times 900$     & 61    & 400 - 1000    & 14   & \xmark    & \xmark    & \xmark & \xmark \\
		Hordley\cite{hordley_multi-spectral_2004}       & $\sim 336 \times 271$    & 31    & 400 - 700     & 23   & \xmark    & \xmark    & \xmark & \xmark \\
		Le Moan\cite{larabi_database_2015}              & $500 \times 500$      & 160   & 410 - 1000    & 9    & \xmark    & \xmark    & \xmark & \xmark \\
		Foster\cite{foster_frequency_2006}              & $1018 \times 1339$    & 33    & 400 - 720     & 8    & \xmark    & \xmark    & \xmark & \xmark \\
		Chakrabarti\cite{chakrabarti_statistics_2011}   & $1392 \times 1040$    & 31    & 420 - 720     & 50   & \xmark    & \xmark    & \xmark & \xmark \\
		BGU\cite{leibe_sparse_2016}                     & $1392 \times 1300$    & 244   & 400 - 1000    & 201  & \xmark    & \xmark    & \xmark & \xmark \\
		Mian\cite{mian_hyperspectral_2012}              & $752 \times 480$      & 33    & 400 - 720     & 31   & \cmark    & \xmark    & \xmark & \xmark \\
		Proposed                               & $1600 \times 1200$    & 31    & 400 - 700     & 1890 & \cmark    & \cmark    & \cmark & \cmark \\
	\end{tabular}
	\vspace*{-0.3cm}
\end{table*}

There are several techniques to record a hyperspectral image.
The essential challenge is to record a 3D datacube using only 2D sensors.
This hyperspectral 3D datacube is depicted in \fig\ref{fig:hs_cube}.
One can imagine combining different layers or single data elements to fit onto a 2D sensor.
Some of the following techniques drop some of these data elements and need a reconstruction process afterwards.
The first class of recording techniques are scanning devices.
Here, one dimension of this 3D datacube is unfolded across the time dimension.
For example, pushbroom sensors\cite{gomez-chova_correction_2008} are scanning one spatial dimension and the spectral dimension using a 2D sensor.
The second spatial dimension is scanned over time by moving the scene or camera accordingly.
Another example of scanning techniques are filter wheels\cite{koenig_practice_1998}, where the spectral dimension is scanned by spinning a wheel, such that different filters are in front of the sensor.
Thus, this technique unfolds the spectral dimension over time.
Obviously, these techniques are not able to capture hyperspectral videos, since the 3D datacube is scanned over time and thus not fully available at each time instance.

\begin{figure}
	\centering

	\includegraphics[]{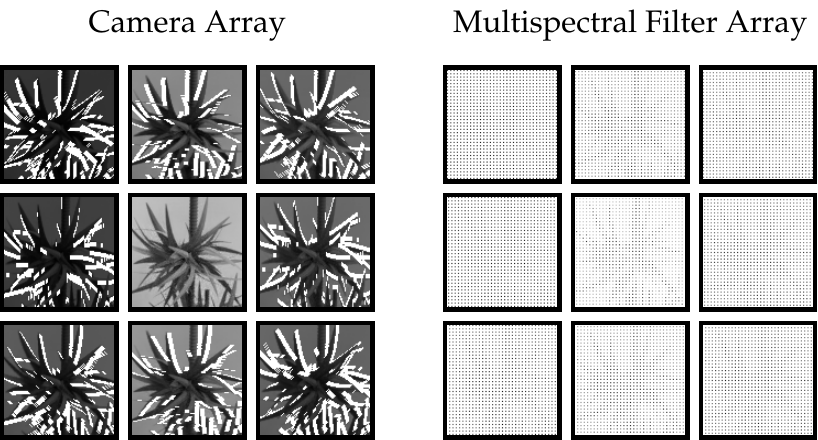}
	
	\caption{Two cases of missing ground-truth pixels of snapshot $3 \times 3$ multispectral/hyperspectral cameras. On the left, the warped channels of a camera array are depicted, where pixels are missing due to occlusions. On the right, the channels of a multispectral filter array sensor are shown. Missing pixels are shown in white.}
	\label{fig:motivation}
\end{figure}

Hyperspectral snapshot cameras, on the other hand, are able to capture hyperspectral videos.
Several ideas exist how to build a snapshot hyperspectral camera.
There are concepts using dispersion and fiber bundles\cite{gat_development_2006} or a 2D dispersion pattern within a computed tomography framework\cite{descour_demonstration_1997}.
Thus, a reconstruction process is mandatory.
Both of these techniques suffer from a very low spatial resolution.
Another technique is multispectral beamsplitting\cite{matchett_volume_2007}, where the incoming light beam is split into the corresponding bands using special mirrors.
The problem here is, that the amount of photons for other bands is also drastically reduced by each mirror.
For a large number of bands, this technique does not work anymore.
Multispectral filter arrays\cite{monno_practical_2015} have a similar idea as a classical Bayer pattern by placing filters periodically over each pixel of a sensor.
Subsequently, a demosaicing algorithm is necessary to yield a full hyperspectral image.
However, depending on the number of bands, the underlying physical resolution for each channel drops significantly using multispectral filter arrays.
Especially, when thinking of hyperspectral imaging which can have up to several hundred bands, the pixels with the same filter are positioned far apart.
Multi-aperture filtered cameras\cite{shogenji_multispectral_2004} have a similar idea, but place filters over whole regions of the sensor.
Thus, the filters are not periodically repeated.
Here, also a reconstruction process is necessary.
Imaging techniques based on compressed sensing\cite{gehm_single-shot_2007} are also discussed.
For these cameras, a coded aperture followed by a dispersion element is used to record different spectral mixtures of different pixels.
Afterwards, compressed sensing is used to reconstruct the datacube.
Finally, works on camera arrays\cite{genser_camera_2020} are currently active as well.
Since the cameras are spatially distributed, a reconstruction pipeline yielding a consistent datacube is essential.
Two examples for the necessity of a reconstruction pipeline for high-resolution hyperspectral snapshot cameras are shown in \fig\ref{fig:motivation}.

From this short overview of hyperspectral cameras, it becomes clear that it is challenging to get ground-truth high-resolution hyperspectral video data.
Ground-truth data has several requirements.
First, the scene itself should be as realistic as possible.
Of course, when recording a scene in reality, this is fulfilled by definition, while it is more of a challenge for synthetic data.
For synthetic data, the world, including all physical concepts like light, needs to be modeled properly to get as realistic data as possible.
Second, ground-truth data needs a high resolution in every dimension to be able to sample in any direction.
While scanning techniques can deliver this as long as the scene is static, some hyperspectral snapshot cameras either record a very low spatial or spectral resolution or have  low underlying spatial or spectral resolution.
Third, the amount of artefacts and postprocessing should be as low as possible.
These artefacts include noise, blur, reconstruction errors, exposure problems, chromatic aberrations and many more.
Here, recording a scene using existing hyperspectral imaging techniques fails to fulfill this requirement.
For a non-moving scene a near ground-truth hyperspectral image can be generated using scanning techniques.
However, for example for a filter wheel, due to different behaviours of lens-filter combinations, the focus and exposure time have to be adjusted, which leads to different behaviours regarding optics and noise.
Snapshot hyperspectral cameras most often need a reconstruction process, which cannot recover the perfect datacube.
Thus, again artefacts may be introduced.
Fourth, boundary conditions should be given as exact as possible.
This includes the used camera(s) including all specifications like sensor size and resolution, as well as the aperture and focal length of the lens.
Furthermore, properties of the bandpass filters and light are also very important.
Since there are always slight deviations during manufacturing processes and aging phenomena, it is impossible to properly fulfill this requirement when recording a scene hyperspectrally in reality.

As one can see, there is a strong need for ground-truth hyperspectral video data, for example also as reference data to measure the performance of snapshot reconstruction algorithms.
There are a number of successful hyperspectral image databases, which are summarized in Table \ref{tab:databases}.
One of the most famous examples is the CAVE database\cite{yasuma_generalized_2010}.
This database contains 32 scenes depicting different objects like balloons, persons,  pepper and color patches usually with a black background.
The images have a resolution of $512 \times 512$ pixels in the range from 400 nm to 700 nm in 10 nm steps, resulting in 31 hyperspectral channels.
The CAVE database does not contain any natural scene.
The UGR database\cite{eckhard_outdoor_2015} contains 14 images with a resolution of $1000 \times 900$ pixels.
This database depicts mostly urban scenes including trees, cars and buildings.
The scenes were recorded from 400 nm to 1000 nm in 10 nm steps, resulting in 61 hyperspectral channels.
Unfortunately, some of the bands are corrupted by artefacts.
The database of Hordley et al.\cite{hordley_multi-spectral_2004} contains 23 images of packagings and color charts and thus does not hold any natural spectra.
The resolution of the images differ.
The packagings and color charts were recorded from 400 nm to 700 nm in 10 nm steps, resulting in 31 hyperspectral channels.
Le Moan et al.\cite{larabi_database_2015} created a database of 9 different textures including textile, wood and skin from 410 nm to 1000 nm in 160 hyperspectral channels.
The textures have a resolution of $500 \times 500$.
The next database of Foster et al.\cite{foster_frequency_2006} contains 8 hyperspectral images of natural and urban scenes.
These were recorded in the range from 400 nm to 720 nm in 10 nm steps, resulting in 33 hyperspectral channels.
The images have a resolution of $1018 \times 1339$.
Chakrabarti et al.\cite{chakrabarti_statistics_2011} recorded a hyperspectral image database of 50 hyperspectral images with a resolution of $1392 \times 1040$.
For each scene, 31 hyperspectral channels were recorded from 420 nm to 720 nm in 10 nm steps.
This database mainly contains urban and indoor scenes.
The BGU database\cite{leibe_sparse_2016} of Arad et al. contains over 200 hyperspectral images of natural, urban and indoor scenes at a resolution of $1392 \times 1300$ pixels.
The database was recorded from 400 nm to 1000 nm at 1.25 nm steps.
Thus, the BGU database offers the highest spectral resolution, one of the highest spatial resolutions and the most diversity in scenes due to the sheer amount of different scenes.
Finally, a hyperspectral video database\cite{mian_hyperspectral_2012} containing a single video of a lab scene with a spatial resolution of $752 \times 480$ exists.
Here, the spectrum from 400 nm to 720 nm was recorded using 10 nm steps, thus resulting in 33 hyperspectral channels.

This paper introduces a novel synthetic hyperspectral video database rendered using a camera array.
This camera array consists of nine cameras arranged in a three times three grid.
Thus, the scenes are rendered from multiple positions for wavelengths between 400 nm and 700 nm in 10 nm steps, resulting in 31 hyperspectral channels for each camera.
Moreover, the depth map is generated as well for every frame and camera.
Since a high resolution representation of spatial positions in three dimensions as well as the spectral dimension is available, many applications can be addressed by this database.
These applications do not necessarily have to be in a hyperspectral context, but also range to grayscale and RGB image processing tasks as well as fusion problems, especially regarding depth sensors.
To the best of our knowledge, this paper is the first to introduce a \textit{synthetic} hyperspectral \textit{video} database.

The remainder of this paper is organized as follows.
First, the techniques and properties of this hyperspectral video database are described in detail in Section \ref{sec:database}.
Afterwards, two applications of this database are considered in Section \ref{sec:applications} by improving the corresponding state of the art.
Finally, the paper is summarized in Section \ref{sec:conclusion}.

\section{Database}
\label{sec:database}
The hyperspectral video database was rendered using the open-source software Blender\cite{blender}, which provides tools for RGB rendering.
Therefore, the first properties and techniques that are described in this section are related to Blender.
The basic idea is to render each wavelength for each frame image by image in grayscale.
For that, the textures need to be adjusted to the wavelength as well as the light source.
Of course, the scene can be rendered from multiple cameras within a camera array.

\subsection{Renderer}

\begin{figure*}[t]
	\centering
	\includegraphics[]{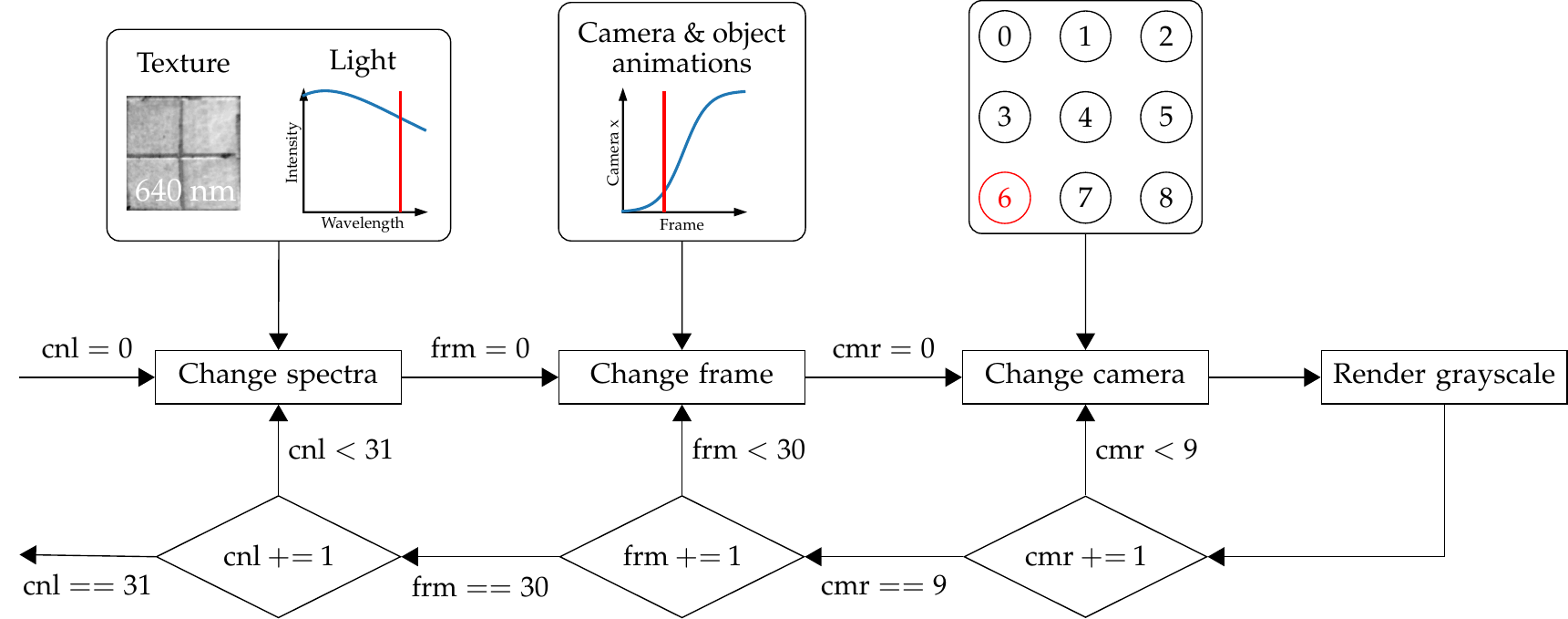}
	
	\caption{The pipeline of the hyperspectral renderer. Essentially, all combinations of channel (cnl), frame (frm) and camera (cmr) are traversed and rendered. The red items in the top row indicate an exemplary state, i.e., currently the textures at 640 nm and the animations at frame 10 of 30 viewed from camera 6 are rendered.}
	\label{fig:rendering_pipeline}
\end{figure*}

To extend Blender to a hyperspectral image renderer, each wavelength is rendered as grayscale image.
Since the wavelengths from 400 nm to 700 nm were rendered in 10 nm steps, one hyperspectral image consists of 31 rendered grayscale images.
For each wavelength, the images of materials as well as the intensity of the light source were changed programmatically through Python.
More details on textures and light sources will be presented below.
This pipeline is explained in detail in \fig\ref{fig:rendering_pipeline}.

Blender has two different rendering engines, namely, Eevee and Cycles.
While Eevee is a simple rasterization engine, Cycles implements path tracing\cite{purcell_ray_2005} and thus yields much more realistic images.
Consequently, Cycles was chosen as rendering engine with branched path tracing as integrator.
The number of samples per pixel was set to 128.

Since path tracing nearly always yields rendering noise in complex scenes, a post-processing denoiser was employed.
Considering that Blender directly implements Non-Local Means (NLM) denoising\cite{buades_non-local_2011}, NLM was used as denoiser after the grayscale images were rendered.

The bitdepth of the final output was set to eight to allow for a smaller total database size.
Furthermore, each grayscale image is saved as PNG.

\subsection{Hyperspectral Textures and Materials}

An essential part for setting up a realistic scene in Blender are textures to give surfaces of objects color and details without increasing the complexity of geometry.
To give a surface the desired part of a texture, coordinates in images are mapped to coordinates of surfaces.
Since a hyperspectral scene also needs hyperspectral textures, the possibilities are either to record them, which requires a high resolution hyperspectral camera, or extract them from other databases.
The latter was picked for this database.
To maintain a consistent appearance of the textures, the BGU database\cite{leibe_sparse_2016} was chosen to deliver hyperspectral textures.
Though this database contains wavelengths from 400 nm to 1000 nm in 1.25 nm increments, our novel database only contains wavelengths from 400 nm to 700 nm in 10 nm steps, to keep the rendering time feasible and the database size acceptable.
The BGU database contains over 200 images of diverse natural, urban, and indoor scenes.
Thus, different textures like grass, wood, flower, leaves, metal, plastic, different types of stones, parts of cars and many more could be extracted from this database.
The sky was also extracted from this database using the image \textit{hill\_0325-1228}, which shows a huge portion of a cloud-free sky.
This texture was then used for the world material in Blender.
Since Blender is only an RGB renderer natively, hyperspectral textures consist of one grayscale image for each wavelength.
Thus, depending on the wavelength that is currently rendered, the textures need to be exchanged.

The scenes have 14 to 34 different textures depending on the complexity.
For example, an outdoor scene needs rather less texture, since there is mainly dirt, grass, wood and leaves.
Then, for these types of materials, different textures for leaves and wood are extracted from the BGU database.
On the other hand, indoor scenes have typically more textures, since there are more smaller things lying around the scene, which all need a texture.
Of course, one can reuse some textures for some objects, but an interesting indoor scene also needs diverse textures even for the same object.

\begin{figure}
	\centering
	\includegraphics[]{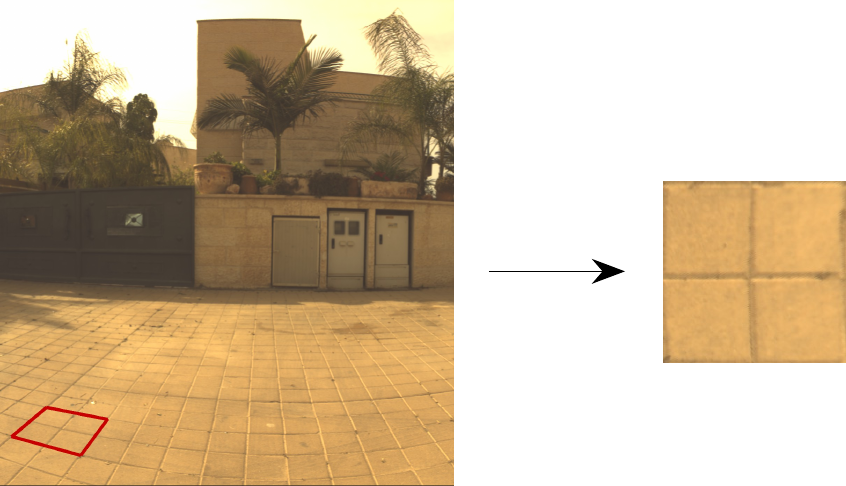}
	\caption{The procedure of extracting hyperspectral textures. The RGB image is taken to spot good textures. The extraction is done by warping all hyperspectral channels using a calculated homography matrix.}
	\label{fig:hte}
\end{figure}

The extraction procedure is shown in \fig\ref{fig:hte}.
First, four points spanning a rectangle are selected.
Afterwards, a homography matrix is calculated, which translates these world coordinates in the image plane to coordinates $\left( (0, 0), (0, \Height), (\Width, \Height), (\Width, 0) \right)$, where $\Width$ and $\Height$ represent the width and the height of the resulting texture.
This has to be defined by the user.
From these four pairs of points $(\xisrc, \yisrc) \leftrightarrow (\xidst, \yidst)$, the homography matrix
\begin{equation}
	\Homography =
	\begin{bmatrix}
		h_{11} & h_{12} & h_{13}\\
		h_{21} & h_{22} & h_{23}\\
		h_{31} & h_{32} & h_{33}
	\end{bmatrix}
\end{equation}

is found by solving the least-squares problem

\begin{equation}
	\begin{split}
		\argmin_{\Homography} \sum_{i=1}^{4} & \left( \xisrc - \frac{h_{11}\yidst + h_{12}\xidst + h_{13}}{h_{31}\xidst + h_{32}\yidst + h_{33}} \right) +\\
		& \left( \yisrc - \frac{h_{21}\yidst + h_{22}\xidst + h_{23}}{h_{31}\xidst + h_{32}\yidst + h_{33}} \right).
	\end{split}
\end{equation}

Since a homography matrix has eight degrees of freedom, the final matrix is scaled such that $h_{33} = 1$.
Then, the position of other points in the destination image plane can be calculated by
\begin{equation}
	\begin{bmatrix}
		\xdst\\
		\ydst\\
		1
	\end{bmatrix} = \Homography
	\begin{bmatrix}
		\xsrc\\
		\ysrc\\
		1
	\end{bmatrix}.
\end{equation}
This is necessary to warp the texture from the source image plane to the destination image texture plane.
This warping procedure is done for every hyperspectral channel.
Afterwards, the hyperspectral channels are saved individually as lossless grayscale images.

After the textures are extracted, corresponding materials can be set up in Blender.
The texture of the material is changed depending on which channel is currently rendered.
Moreover, a lot of properties can be changed in Blender, which influences how light is reflected or absorbed.
For example, the roughness can be changed as well as how specular and metallic a material is, e.g., a shiny car paint is typically very little rough but very metallic, while a street is very rough and less shiny.
Moreover, also the transparency of a texture can be set, which is very important for windows.

\subsection{Light Sources}

\begin{figure}
	\centering
	\includegraphics[]{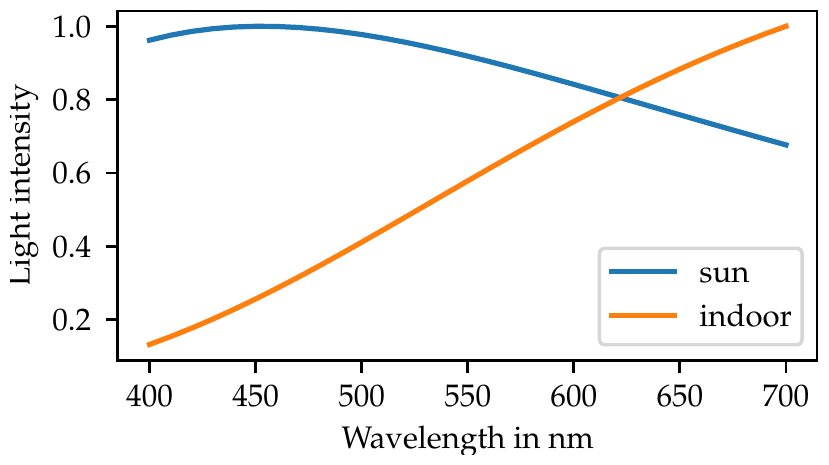}
	\caption{Used light spectra for the sun (6400K) and the indoor lamp (3200K).}
	\label{fig:light_spectra}
\end{figure}

For all outdoor scenes or scenes with windows, the light source is the sun.
In bright daylight the sun can reach a color temperature up to 6400K.
The color temperature of a light source defines its emitting light spectrum.
The spectrum is based on the radiation of an ideal black body and can be calculated by
Planck's law.
For the single indoor scene without windows, a lamp was mounted into that room with a color temperature of 3200K, which is a typical value for studio lamps.
The spectra of these two light sources is shown in \fig\ref{fig:light_spectra} and were calculated by Planck's law.
For each wavelength, the intensity of the sun or the indoor lamp in the scene is set according to this plot.
Furthermore, the hardness and light strength differs from scene to scene and is also dependent on the angle of the sun.

\subsection{Cameras and Camera Array}

\begin{figure}[t]
	\centering
	\includegraphics[]{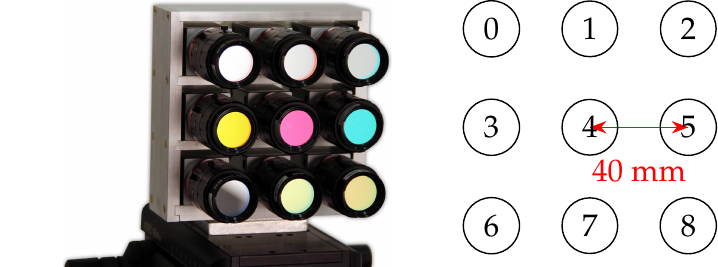}
	\caption{A multispectral camera with nine channels on the left\cite{genser_camera_2020} with a sketch of dimensions and indices of cameras on the right. Note that this view is from the front, as well.}
	\label{fig:camsi}
\end{figure}

All scenes are rendered from nine cameras arranged in a $3 \times 3$ grid.
This mimicks a camera array from Genser et al.\cite{genser_camera_2020} and is shown in \fig\ref{fig:camsi}.
Therefore, the specifications are very similar.
The baseline between neighboring cameras is 40 mm.
The cameras itself are very close to common industrial cameras like a Basler acA1600-60gm, which are also used for the camera array in \fig\ref{fig:camsi}.
Thus, the image sensor size is set to 7.2 mm times 5.4 mm and the resolution of the images is 1600 pixels times 1200 pixels.
The focal length is different from the original camera array and is set to 6 mm focal length to capture a large portion of the scene.
The purpose of rendering the scene from this camera array is, for example, to be able to simulate the multispectral camera shown in \fig\ref{fig:camsi}.
Since this multispectral camera array only records one spectral area with a specific transmission curve per camera, one has to simulate the corresponding filters by calculating a weighted sum of hyperspectral images for each camera.
Thus, in contrast to the camera array in Genser et al.\cite{genser_camera_2020}, the synthetic data contains all 31 hyperspectral channels for every camera.
If the application just demands for a single hyperspectral view of the scene, only the center camera can be extracted, since each camera has its own folder in the database.

\subsection{Depth Maps}

Many applications also rely on depth data provided by a depth sensor like RADARs or LIDARs.
Moreover, some algorithms also estimate the depth as intermediate step, for example, to register multiple views on top of each other.
Therefore, depth data for all scenes, cameras, and frames is provided.
The depth maps are stored in the EXR format as 32 bit floats for a high resolution.
The depth maps contain the physical depth in meters.
Consequently, the disparity between cameras is calculated by

\begin{equation}
	\label{eq:depth_to_disp}
	\disparity = \frac{\baseline \cdot \focallength}{\depth \cdot \pixelsize},
\end{equation}
where $\disparity$ is the disparity of corresponding pixels between two cameras, $\baseline$ is the baseline between the cameras, e.g. 40 mm for directly neighboring cameras in the camera array, $\focallength$ is the focal length, thus 6 mm, $\depth$ is the depth itself, and $\pixelsize$ is the size of a single pixel and calculates to $\pixelsize = 7.2 \ \text{mm}/1600 = 5.4 \ \text{mm}/1200 = 4.5 \cdot 10^{-3} \ \text{mm}$.
Note that the depth is provided in meters while all other properties are given in millimeters.
Depth maps can be used for example to evaluate depth estimation algorithms or to warp different views of the camera array to different positions.

\subsection{Motion}

In all of the scenes, the camera array moves along a specified path.
Depending on the scene, this is realized using different approaches.
In the first method, the camera array is moved along the scene manually and this trajectory is recorded.
After recording this movement, the paths are smoothed out, since a manual camera movement is typically quite shaky.
In the end, each frame is a keyframe for the movement and stores the position and rotation of the camera array.
The second way to define movement in this database is to define the starting and endpoint position and rotation and smoothly interpolation between them for intermediate frames.
Thus, there are only two keyframes set and the intermediate points are determined using Bezier interpolation.
This approach is also performed for moving objects.

\subsection{Scenes}

\begin{figure}
	\centering
	\includegraphics[]{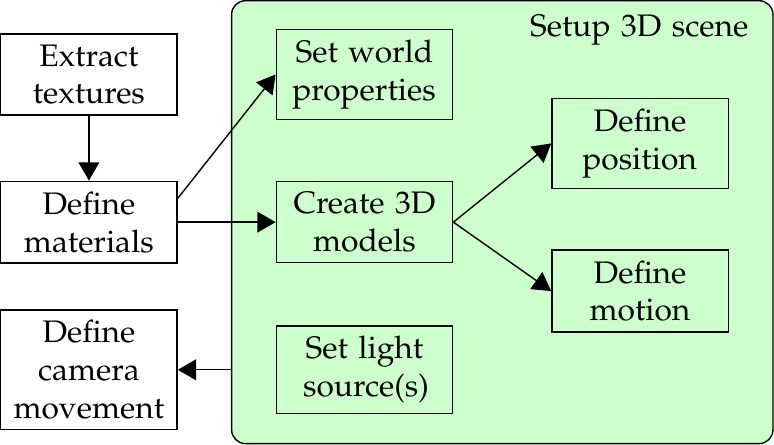}
	
	\caption{The workflow to create a hyperspectral scene. This general workflow can be executed in any rendering framework.}
	\label{fig:workflow}
\end{figure}

\begin{figure*}
	\centering
	\includegraphics[]{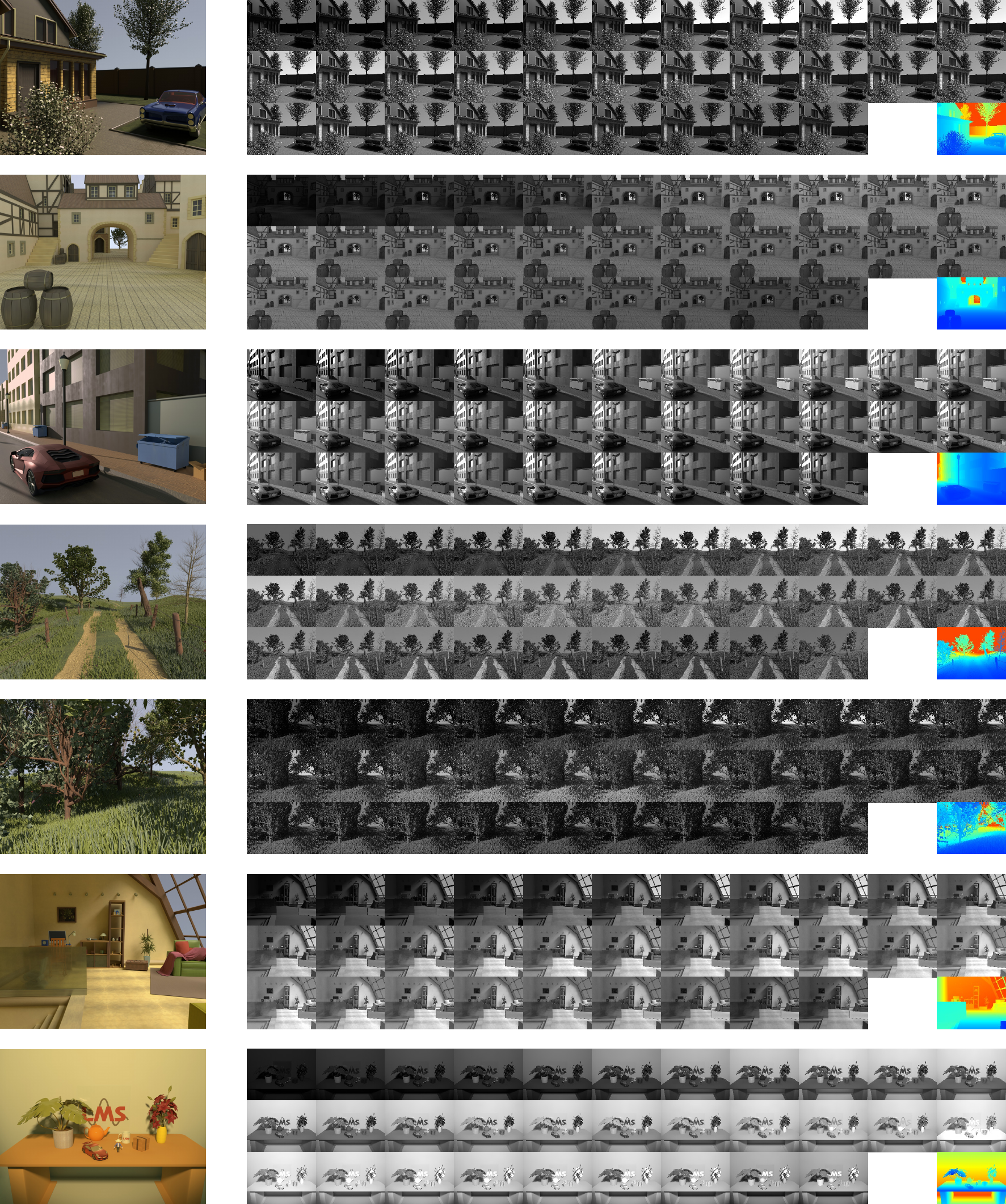}
	\caption{The center views of the seven different synthetic hyperspectral scenes: frame 0 of "family house", frame 0 of "medieval seaport", frame 25 of "city", frame 0 of "outdoor", frame 10 of "forest", frame 0 of "indoor" and frame 14 of "lab". On the left, an RGB image using standard CIE 1931 color curves can be seen. Next to it, all 31 hyperspectral channels of the corresponding image and the depth map are depicted.}
	\label{fig:scenes}
\end{figure*}

The whole workflow for setting up a 3D hyperspectral scene is summarized in \fig\ref{fig:workflow}.
In total, seven different scenes were rendered.
These scenes are depicted in \fig\ref{fig:scenes}, where the RGB images are calculated using standard CIE 1931 color curves.
Furthermore, all 31 hyperspectral channels of one frame are shown as well as the depth maps corresponding to the respective scene and frame.
The camera array moves through each scene for 30 frames.

There are five outdoor scenes. \textit{Family house} represents a mixture of urban and rural objects, \textit{medieval seaport} is a view of a more ancient city incorporating a lot of stone and wood, and \textit{city} is a newer city including a moving car, normal streets, sidewalks, modern houses and trash bins.
\textit{Outdoor} represents a rural scene from an agricultural way and \textit{forest} is a similar scene without any man-made objects and a lot of trees.
\textit{Indoor} depicts a room including furniture and decoration elements, but still with the light source being the sun shining through windows.
Finally, \textit{lab} is illuminated by a halogen lamp and depicts a laboratory setup with selective elements on a table similar to the real-world image taken in Genser et al.\cite{genser_camera_2020}.

\subsection{Limitations}

Note that this database still contains only \textit{synthetic} hyperspectral data.
Thus, it is nearly clean from any noise or artefacts and ideal cameras are assumed as well as an ideal camera array is assumed.
An array-related real-world artefact would be that the cameras are not perfectly aligned to each other as well as the sensors within the camera housing are not perfectly fitted.
Therefore, in real-world a calibration procedure between different cameras would always be necessary, e.g., to fulfill the epipolar constraint.
Moreover, a lot of artefacts of the filter-lens-camera combinations can occur.
First, filters have the tendency to change optical paths slightly dependent on the wavelength.
Second, lenses introduce artefacts like optical veiling glare\cite{mccann_veiling_2007}, lens flares\cite{rosenhauer_image_1968}, wavelength-dependent chromatic aberration\cite{waller_phase_2010}, star bursts and many more.
Finally, cameras always introduce different kind of noise, e.g., shot noise, thermal noise and electronic noise\cite{hytti_noise_2006}.
All of these artefacts are not part of this database and most of them are difficult to simulate.
Thus, if one trains or evaluates algorithms based on this data, one has to be careful, since the data of this database deviates from real-world image acquisitions.
However, depending on the desired application, noise or artefacts could be added to clean data.
Moreover, while the absolute performance is wrongly estimated due to missing artefacts, the relative performance between two algorithms might still be possible assuming that both algorithms are affected by noise and artefacts equally.

\section{Applications}
\label{sec:applications}
In this section, an overview over applications that can benefit from this dataset is given.
First, two applications are presented in detail and improved which is verified using the novel database.
In the first application, videos from all cameras and depth maps from the center camera are used.
The second application uses just the hyperspectral video of the center camera.
In the end, several other applications are described briefly.

\subsection{Cross Spectral Reconstruction}

\begin{figure}
	\centering
	\includegraphics[]{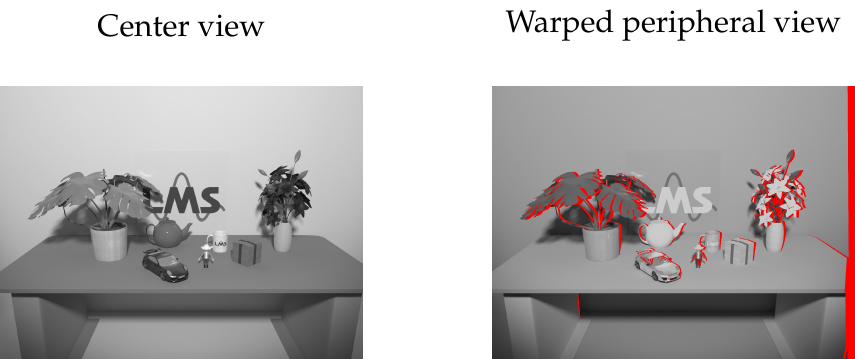}
	\caption{The basic problem of cross spectral reconstruction. On the left side the fully available center view is depicted, on the right side the warped peripheral view is shown. Due to occlusions, some pixels are missing in the peripheral view, which are shown in red.}
	\label{fig:CSR_problem}
\end{figure}

The first application that is covered in detail is cross-spectral image reconstruction.
Consider imaging a scene with the camera array shown in \fig\ref{fig:camsi}.
The first step to create a consistent multispectral datacube is a cross-spectral disparity estimation, i.e., finding the correspondence of which pixel in one camera belongs to which pixel in the other camera.
Since the final reconstruction process is evaluated here, the ground-truth depth maps from the database can be used.
For that, the depths needs to be converted to disparities using \eqref{eq:depth_to_disp}.
Then, this disparity map can be used to warp the peripheral view to the center view.
Due to occlusions, there are pixels missing which need to be reconstructed.
Fortunately, the center view is always fully available, since it is not warped to any other position.
Thus, the center view can be used as reference channel for all peripheral views.
However, this center view records a different part of the spectrum than the peripheral views.
Note that the procedure to simulate a multispectral camera array is described in Section \ref{subsec:csr_eval}.
Therefore, one has to find a relationship between the center view and the peripheral view that has to be reconstructed.
This whole problem is depicted in \fig\ref{fig:CSR_problem}.
The peripheral view that has to be reconstructed will be called distorted view in the following description.

\subsubsection{State of the Art}

For a single multispectral image, Non-local Cross-Spectral Reconstruction (NOCS) in \cite{sippel_spatio-spectral_2021} is a state-of-the-art approach.
NOCS first finds similar blocks using the fully available reference images.
Block means the set of pixels of usually a square area around a center pixel.
In the case of the multispectral camera array, this would be just the center view.
For that, the $l_2$-norm between two blocks within the reference image is minimized
\begin{equation}
	d_{\text{NOCS}}(\x, \y) = ||\Block^{\text{R}}(\x) - \Block^{\text{R}}(\y)||_2,
\end{equation}
where $\Block^{R}(\x)$ describes the reference block at coordinate $\x$.
Of course, this whole block matching procedure only has to be done for pixels that are missing in the distorted view.
For missing pixel position $\x$, the $\BestMatches$ best matches are used to stack the corresponding pixel values of the reference view on top of each other, resulting in the vector $\ReferenceVector(\x)$.
The same procedure is done for the distorted image, resulting in the distorted vector $\DistortedVector(\x)$.
Of course, this vector contains missing pixels.
Moreover, since a block has distance 0 with itself, the pixel to reconstruct is always at the first entry of the vector.

These two vectors are then used to build a linear regression model
\begin{equation}
	\DistortedVector(\x) = \LinRegA(\x) \cdot \ReferenceVector(\x) + \LinRegB(\x),
\end{equation}
where $\LinRegA(\x)$ and $\LinRegB(\x)$ are the parameters of the linear regression and thus the variables that need to be estimated for every missing pixel.
They are found by minimizing the $l_2$-norm of the difference between known references of the distorted vector $\tilde{\DistortedVector}(\x)$ and its prediction from the reference vector $\tilde{\ReferenceVector}(\x)$ at the corresponding positions
\begin{equation}
	\hat{\LinRegA}(\x), \hat{\LinRegB}(\x) = \argmin_{\LinRegA(\x), \LinRegB(\x)} ||\LinRegA(\x) \cdot \tilde{\ReferenceVector}(\x) + \LinRegB(\x) - \tilde{\DistortedVector}(\x)||_2^2.
\end{equation}
These parameters are found in closed-form\cite{sippel_spatio-spectral_2021}.
Afterwards, this model is used to predict the missing pixel, which is the first element in the distorted vector.
For that, the value of the first element in the reference vector is put into the model and the resulting value is the predicted reconstructed pixel value
\begin{equation}
	\DistortedVector_1(\x) = \hat{\LinRegA}(\x) \cdot \ReferenceVector_1(\x) + \hat{\LinRegB}(\x).
\end{equation}

Of course, not all missing pixels are reconstructed at once, but an iterative procedure is carried out.
Always 10\% (but at least one) of the remaining pixels to reconstruct that have the highest number of non-missing pixels of the distorted view in their vector are reconstructed.
Afterwards, all vectors that contain a newly reconstructed pixel are updated and the loop starts from the beginning.
Thus, reconstructed pixels will influence the remaining pixels that need to be reconstructed.

\subsubsection{Novel Approach}

We propose to extend this method by also using the previous frame and therefore exploit temporal correlation to enhance NOCS.
This novel method is called Temporal Cross Spectral Reconstruction (TNOCS).
To exploit temporal correlation, the block matching procedure is also executed on the previous frame.
Therefore, the block matching procedure needs to consider that similar blocks might be in different frames
\begin{equation}
	d_{\text{TNOCS}}(\x, \y, t_d) = ||\Block^{\text{R}}_{t}(\x) - \Block^{\text{R}}_{t - t_d}(\y)||_2,
\end{equation}
where $\Block^{R}_{t}(\x)$ is a block in the reference video at spatial position $\x$ and current frame $t$, and $t_d$ is the difference to the respective frame.
Therefore, when a similar block for $\x$ is searched, the position and the frame of the other block can be varied.
Since two consecutive frames are usually highly correlated but still depict the scene from a slightly different angle, only the previous and the current frame are searched for similar blocks.
Thus, $t_d$ can only be in the set of $\{0, 1\}$.
The remaining formulas are the same, since the pixel at position $\x$ still needs to be reconstructed and the vectors contain pairs of reference pixels and the corresponding pixels from the distorted video.

\begin{figure}
	\centering
	\includegraphics[]{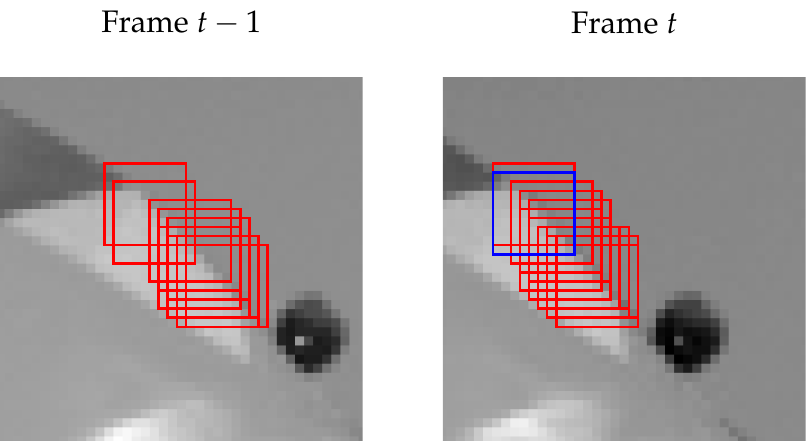}
	
	\caption{The proposed TNOCS searches similar blocks in the current frame as well as the previous frame to build the linear regression model. The blue box depicts the block for which similar blocks are searched, the red blocks are the best matches. The images depict cutouts of two consecutive frames of the \textit{lab} scene.}
	\label{fig:TNOCS}
\end{figure}

For an improvement over NOCS, it is essential that the camera or object in scene moves from one frame to another frame, since otherwise the pixels to be reconstructed would be exactly the same and the temporal correlation can not be properly exploited.
Note that only non-reconstructed pixels of the previous frame are used for block matching, since the error in reconstruction should not propagate through all frames.
Otherwise, this would lead to severe artifacts after some frames.
Of course, the block matching is also executed on the current frame to further exploit spatial correlation.
This idea is summarized in the \fig\ref{fig:TNOCS}.

\subsubsection{Evaluation}
\label{subsec:csr_eval}

To validate that this idea works well, our novel hyperspectral video database is employed.
For that, multispectral video data has to be created out of the hyperspectral data.
To achieve this, an imaging pipeline needs to be simulated.
A single pixel of the $i$-th channel of a multispectral image is recorded by\cite{sippel_structure-preserving_2020}

\begin{equation}
	\MSPixel_i = \int \LightSource(\Wavelength)\Reflectance(\Wavelength)\Filter_i(\Wavelength)\CameraLens(\Wavelength)\CameraSensor(\Wavelength) \text{d}\Wavelength,
\end{equation}
where $\LightSource(\Wavelength)$ is the spectrum of the light source, $\Reflectance(\Wavelength)$ is the reflectance spectrum of the material, $\Filter_i(\Wavelength)$ is the transfer function of the $i$-th filter corresponding to the $i$-th multispectral channel, $\CameraLens(\Wavelength)$ the spectral permeability of the camera lens and $\CameraSensor(\Wavelength)$ is the spectral response of the camera sensor.
In this case, the database already yields a sampled version of the product of the light source and reflectance $\HSImageSpectrum(\Wavelength) = \LightSource(\Wavelength)\Reflectance(\Wavelength)$ as images.
The transfer function of the camera lens and camera sensor are assumed to be perfect in the covered wavelength area from 400 nm to 700 nm.
Since the images are obviously sampled in the spectral dimension, this equation needs to be sampled as well.
Therefore, a single hyperspectral pixel from our database can be described by $\HSPixel = [\HSImageSpectrum(\Wavelength_1), \HSImageSpectrum(\Wavelength_2), \dots, \HSImageSpectrum(\Wavelength_{31})]\T$.
When stacking the sampled filters on top of each other into the filter matrix $\FilterMatrix = [\Filter_i(\Wavelength_1), \Filter_i(\Wavelength_2), \dots, \Filter_i(\Wavelength_{31})]$, the imaging process for a single pixel can be written as
\begin{equation}
	\MSPixel = \FilterMatrix \HSPixel.
\end{equation}

In this case, the filter matrix contains sampled values of real filters used for the camera array.
Namely, six bandpass filters with bandwidth 50 nm at wavelengths 425 nm, 450 nm, 500 nm, 550 nm, 600 nm and 650 nm, and a red filter, a green filter and a blue filter to compose an RGB image.
Then, the data to reconstruct is created by calculating the multispectral data for the corresponding camera.
The ground truth is calculated by doing the same calculations with the middle camera.
Moreover, the warped images are calculated using the ground-truth depth maps provided by the database.

\begin{table}[t]
	\centering
	\renewcommand{\arraystretch}{1.5}
	\caption{Evaluation in terms of PSNR of the novel temporal cross spectral reconstruction algorithm (TNOCS) against the version without exploiting temporal correlation (NOCS).}
	\label{tab:eval_reconstruction}
	\begin{tabular}{c|c|c}
		& NOCS      & TNOCS             \\ \hline
		Family house        & 30.83 dB  & \textbf{36.43 dB} \\ \hline
		Medieval seaport    & 40.06 dB  & \textbf{44.87 dB} \\ \hline
		City                & 43.61 dB  & \textbf{45.39 dB} \\ \hline
		Outdoor             & 28.94 dB  & \textbf{34.05 dB} \\ \hline
		Forest              & 27.03 dB  & \textbf{30.67 dB} \\ \hline
		Indoor              & 39.33 dB  & \textbf{41.15 dB} \\ \hline
		Lab                 & 35.40 dB  & \textbf{36.26 dB} \\ \hline \hline
		Average             & 35.03 dB  & \textbf{38.40 dB} \\
	\end{tabular}
	\vspace*{-0.3cm}
\end{table}

Using this multispectral database, the novel TNOCS algorithm can be evaluated against its non-temporal version NOCS.
The results are summarized in Table\ \ref{tab:eval_reconstruction}.
From the table it gets obvious that exploiting temporal correlation by just searching for similar blocks in the previous frame is enhancing the performance significantly.
Especially for scenes containing a lot of leaves, which leads to many small areas that are missing, the difference is enormous.
For other scene like \textit{lab} order \textit{city}, the improvement is smaller but still noticeable.

Without our novel hyperspectral video database, it would be impossible to accurately validate the performance gain of exploiting temporal correlation.
As shown, it is also possible to create nearly any other data that lies in the wavelength area from 400 nm to 700 nm.

\subsection{Hyperspectral Video Coding}

The second application addressed is hyperspectral video coding.
The need for hyperspectral image and video coding is basically already motivated by this database.
With just seven scenes, nine cameras and 30 frames, this database already has a size bigger than 60 GB saved as single grayscale PNGs.
When looking at raw values, this number increases even more.
For this hyperspectral database, saving all images without any compression and eight bits per grayscale pixel results in $7 \cdot 9 \cdot 30 \cdot 31 \cdot 1600 \cdot 1200 \ \text{bytes} \approx 105 \ \text{GB}$.
This enormous amount of data shall be compressed, since there is a lot of correlation.
Of course, there is temporal correlation, which can be exploited.
But also the different hyperspectral channels are highly correlated which can be exploited by a coder.
In the following, the term encoding means producing a compressed bitstream out of a (hyperspectral) video.
In contrast, the term decoding means transforming this bitstream back to the (hyperspectral) video.
For lossy coders, this decoded video deviates from the original video.

\subsubsection{State of the Art}

The presented method is based on a multispectral and hyperspectral image coder named Pel-Recursive Inter-Band Prediction (PRIBP)\cite{meyer_multispectral_2020}.
Before presenting the novel inter component of this coder, the intra hyperspectral coder is reviewed first.
The basic idea of\cite{meyer_multispectral_2020} is to first code three channels, in this paper RGB channels, with a standard coder.
Then, the three decoded channels serve as reference channels for the next channels to code.
The decoded channels need to be used as reference channels, because the other end of the communication channel only has the decoded images, which slightly deviates from the original images depending on the quality settings, available.
Thus, to avoid a drift between encoder and decoder, the original channels cannot be used as reference channels.
For that, more intra modes have been implemented into the High Efficiency Video Coding (HEVC) codec\cite{sullivan_overview_2012}.
The prediction is done on a transform block level using previously predicted pixels and reconstructed pixels from adjacent blocks.
To code the center pixel $\PRIBPBlockPixel$ of the current block $\PRIBPBand$ of the channel to code, first the best reference channel block $\PRIBPRef$ out of the current three reference channels is searched by picking the one with the maximum correlation with the already predicted pixels
\begin{equation}
	\rho(\PRIBPKnownRef, \PRIBPKnownBand) = \frac{\sum_{i=1}^{N} \left( \PRIBPKnownRef_i - \mathbb{E}\{\PRIBPKnownRef\} \right)\left( \PRIBPKnownBand_i - \mathbb{E}\{\PRIBPKnownBand\} \right)}{\sqrt {\sum_{i=1}^{N} \left( \PRIBPKnownRef_i - \mathbb{E}\{\PRIBPKnownRef\} \right)^2} \sqrt{\left( \PRIBPKnownBand_i - \mathbb{E}\{\PRIBPKnownBand\} \right)^2}},
\end{equation}
where $\mathbb{E}\{\cdot\}$ is the expectation operator, $N$ is the number of already predicted pixels, and $\PRIBPKnownRef$ and $\PRIBPKnownBand$ are the already predicted pixels of the current block of the reference signal and the channel, respectively.
Subsequently, this reference is used to build a linear regression model very similar to the one used in cross-spectral reconstruction.
Therefore, the linear model reads as
\begin{equation}
	\PRIBPKnownBand = \LinRegA \cdot \PRIBPKnownRef + \LinRegB
\end{equation}
with parameters $\LinRegA$ and $\LinRegB$ to be estimated.
Again, the parameters are found by minimizing the $l_2$-norm
\begin{equation}
	\hat{\LinRegA}, \hat{\LinRegB} = \argmin_{\LinRegA, \LinRegB} ||\LinRegA \cdot \PRIBPKnownRef + \LinRegB - \PRIBPKnownBand||_2^2.
\end{equation}
Finally, this model is employed for the prediction of the pixel
\begin{equation}
	\hat{\PRIBPBlockPixel} = \hat{\LinRegA} \cdot \PRIBPRefPixel + \hat{\LinRegB}.
\end{equation}
The coding order of the remaining channels is found by sorting the different channels according to their structural similarity index with respect to one anchor channel.
Since there are always three reference channels, the least similar reference channel is then replaced by the channel, which was just coded.
Note that this method was focused on multispectral image coding rather than hyperspectral image coding.

\subsubsection{Novel Approach}

\begin{figure*}
	\centering
	\includegraphics[]{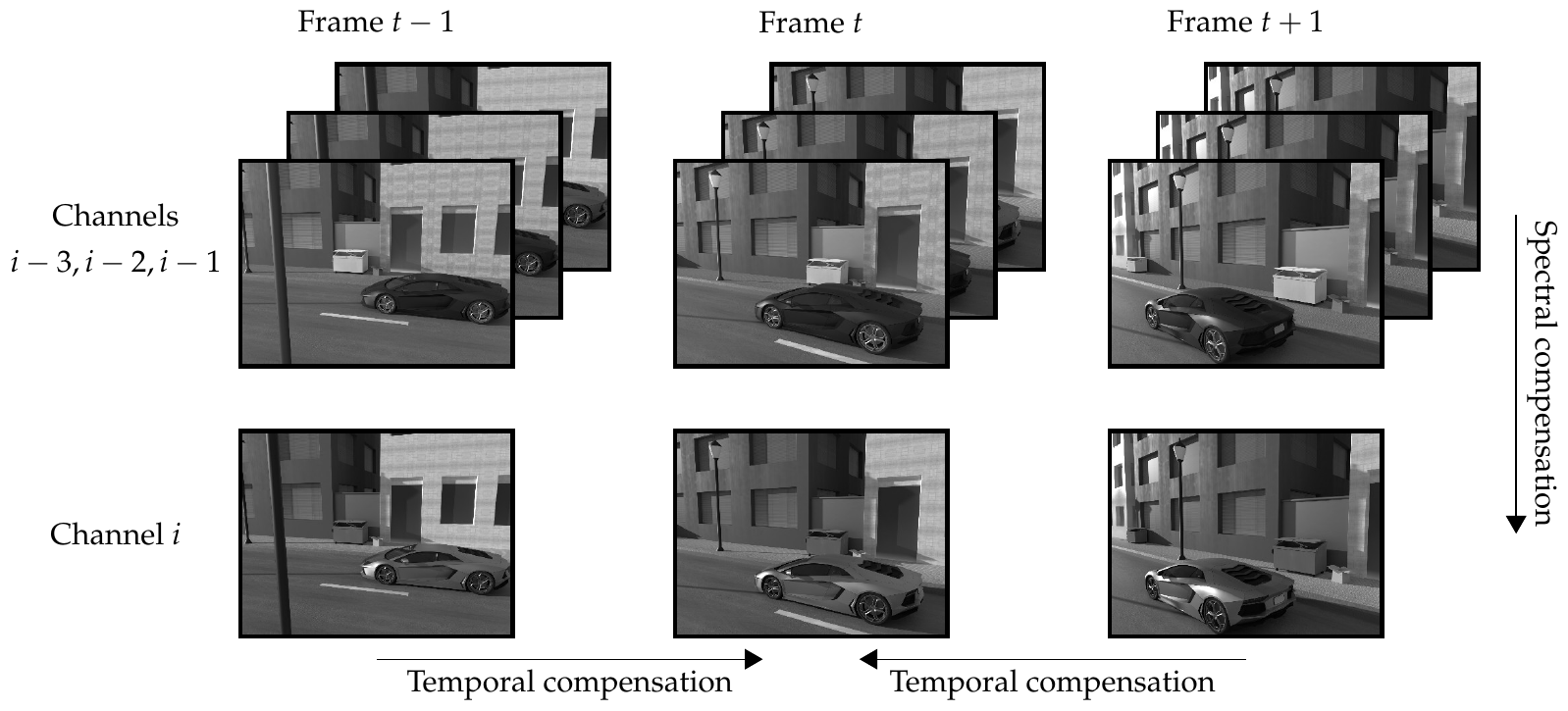}
	
	\caption{The basic concept of this novel inter hyperspectral video coder. First, motion compensation is executed on the frame $t$ exploiting the neighboring frames. Afterwards, the current channel is spectrally compensated using the last three channels.}
	\label{fig:inter_hs}
\end{figure*}

For hyperspectral image coding, there is implicitly more structure given, since neighboring channels are also neighbors in the spectrum.
Therefore, since spectra are typically smooth, the correlation between two neighboring channels is the highest.
The temporal and spectral compensation of the novel hyperspectral video coder is shown in \fig\ref{fig:inter_hs}.
Assuming the channels are sorted in ascending order according to their wavelength, the proposed hyperspectral video coder compresses the first three channels as an RGB video with a standard video coder using only intra-prediction to ensure a high quality for motion estimation using these channels.
Since this work is based on PRIBP, the base video coder is again HEVC.
Thus, the first three decoded channels for every time step $t$ read as
\begin{equation}
	\hat{\InterChannel}^t_0, \hat{\InterChannel}^t_1, \hat{\InterChannel}^t_2 = \InterDecoder^{\text{HEVC}}\left( \InterCoder^{\text{HEVC}} \left( \left[ \InterChannel^t_0, \InterChannel^t_1, \InterChannel^t_2 \right] \right) \right),
\end{equation}
where $\InterCoder^{\text{HEVC}}\left( \cdot \right)$ and $\InterDecoder^{\text{HEVC}}\left( \cdot \right)$ are the coder and decoder, respectively, and $\InterChannel^t_i$ is the $i$-th grayscale channel of a hyperspectral image of size $\InterWidth \times \InterHeight$ at time step $t$.
To avoid issues with error propagations, every second frame of every other channel is coded by the hyperspectral intra encoder, while for all other frames a motion-compensated residual is compressed by the hyperspectral intra coder.
Otherwise, the errors during coding would propagate into the next frame leading to even more errors and a higher residual to code, which would then turn into an even lower quality or higher rate.
This residual is calculated by exploiting temporal correlation.
Therefore, the decoded first three channels are used to estimate motion, which is done by PWC-Net\cite{sun_pwc-net_2018}.
This is done using forward and backward motion estimation
\begin{equation}
	\label{eq:motion_vectors}
	\begin{split}
		\InterMotion^t_{\text{fw}} &= \text{PWC}\left( \left[ \hat{\InterChannel}^{t-1}_0, \hat{\InterChannel}^{t-1}_1, \hat{\InterChannel}^{t-1}_2 \right], \left[ \hat{\InterChannel}^{t}_0, \hat{\InterChannel}^{t}_1, \hat{\InterChannel}^{1}_2 \right] \right) \\
		\InterMotion^t_{\text{bw}} &= \text{PWC}\left( \left[ \hat{\InterChannel}^{t+1}_0, \hat{\InterChannel}^{t+1}_1, \hat{\InterChannel}^{t+1}_2 \right], \left[ \hat{\InterChannel}^{t}_0, \hat{\InterChannel}^{t}_1, \hat{\InterChannel}^{1}_2 \right] \right),
	\end{split}
\end{equation}
where $\InterMotion^t$ is an image of motion vectors for time step $t$.
Then, these motion estimations can be used for motion compensation
\begin{equation}
	\label{eq:motion_compensation}
	\begin{split}
		\InterPrediction^t_{\text{fw}, i}[m, n] &= \tilde{\InterChannel}^{t - 1}_i \left( m + \InterMotion^t_{\text{fw}, 1}[m, n], n + \InterMotion^t_{\text{fw}, 2}[m, n] \right) \\
		\InterPrediction^t_{\text{bw}, i}[m, n] &= \tilde{\InterChannel}^{t + 1}_i \left( m + \InterMotion^t_{\text{bw}, 1}[m, n], n + \InterMotion^t_{\text{bw}, 2}[m, n] \right),
	\end{split}
\end{equation}
where $\tilde{\InterChannel} \left( m, n \right)$ is the interpolated and extended image of $\hat{\InterChannel} \left[ m, n \right]$.
Moreover, corresponding masks, when the motion compensation tries to pull a pixel outside the image scope, need to be calculated
\begin{equation}
	\label{eq:motion_mask}
	\InterMask^t_{j}[m, n] = \left\{
	\begin{array}{ c l }
		1, &\text{if } \begin{aligned}
			&0 \leq m + \InterMotion^t_{j, 1}[m, n] \leq \InterWidth - 1 \ \land \\
			&0 \leq n + \InterMotion^t_{j, 2}[m, n] \leq \InterHeight - 1
		\end{aligned} \\
		0, &\text{else},
	\end{array}
	\right.
\end{equation}
where $j$ is in the set of $\left\{ \text{fw}, \text{bw} \right\}$.
Since two predictions for the same image are made, there has to be a merge procedure for these.
In general, the simple average of these two predictions is taken to merge them.
However, at the border of the image, it occurs that the warping tries to warp pixels outside of the original image plane to the corresponding pixel in the predicted image.
In these situations, the merge procedure entirely relies on the other prediction as long as this prediction is valid.
If no prediction is valid, again the average of both predictions is taken as best guess.
This summarizes to the final prediction
\begin{equation}
	\label{eq:final_prediction}
	\begin{split}
		\InterPrediction^t_i = \left\{
		\begin{array}{c l}
			\frac{\InterPrediction^t_{\text{fw}, i} + \InterPrediction^t_{\text{bw}, i} }{2}, &\begin{aligned}
				&\text{if } \left(\InterMask^t_{\text{fw}}[m, n] = 1 \land \InterMask^t_{\text{bw}}[m, n] = 1\right) \\
				&\lor \left(\InterMask^t_{\text{fw}}[m, n] = 0 \land \InterMask^t_{\text{bw}}[m, n] = 0\right)
			\end{aligned}\\
			\InterPrediction^t_{\text{fw}, i}, &\text{if } \InterMask^t_{\text{fw}}[m, n] = 1 \land \InterMask^t_{\text{bw}}[m, n] = 0 \\
			\InterPrediction^t_{\text{bw}, i}, &\text{if } \InterMask^t_{\text{fw}}[m, n] = 0 \land \InterMask^t_{\text{bw}}[m, n] = 1.
		\end{array}
		\right.
	\end{split}
\end{equation}
To calculate the residual for all odd frames, the predicted image is subtracted from the original decoded image
\begin{equation}
	\label{eq:residual}
	\InterResidual^t_i = \InterChannel^t_i - \InterPrediction^t_i \quad \forall \ t \bmod 2 = 1.
\end{equation}
This yields three residual images which serve as reference residual images for the hyperspectral intra coder, later.
Moreover, as in PRIBP, for all-intra coded frames the first three channels serve as reference channel for the first hyperspectral intra coded image.

Subsequently, all other channels are coded iteratively.
Even frames are coded using all-intra PRIBP
\begin{multline}
	\hat{\InterChannel}^t_i = \InterDecoder^{\text{PRIBP}}\left( \InterCoder^{\text{PRIBP}} \left( \InterChannel^t_i, \left[ \hat{\InterChannel}^t_{i - 1}, \hat{\InterChannel}^t_{i - 2}, \hat{\InterChannel}^t_{i - 3} \right] \right) \right) \\
	\forall \ i \geq 3 \land t \bmod 2 = 0,
\end{multline}
where $\InterCoder^{\text{PRIBP}} \left( \InterChannel^t_i, \left[ \hat{\InterChannel}^t_{i - 1}, \hat{\InterChannel}^t_{i - 2}, \hat{\InterChannel}^t_{i - 3} \right] \right)$ means that the $i$-th channel is coded using decoded channels $i - 1$, $i -2$ and $i - 3$ as reference channels.
After being decoded, they replace the oldest reference in the reference buffer for that frame and serve as reference channel for the next three channels.
This is done, since, as already described, spectra are typically smooth and therefore, the last three channels have the highest correlation with the current channel from a physical point of view.
Then, the decoded intra frames at timesteps $t - 1$ and $t + 1$ are used for predicting the current frame at time $t$ using the motion vectors estimated from the first three channels in \eqref{eq:motion_vectors}.
The motion-compensated prediction is calculated using \eqref{eq:motion_compensation}, \eqref{eq:motion_mask} and \eqref{eq:final_prediction}.
Afterwards, the residual with the frame to code is calculated by \eqref{eq:residual}.
This residual is fed into PRIBP, which uses three reference images again
\begin{multline}
	\hat{\InterResidual}^t_i = \InterDecoder^{\text{PRIBP}}\left( \InterCoder^{\text{PRIBP}} \left( \InterResidual^t_i, \left[ \hat{\InterResidual}^t_{i - 1}, \hat{\InterResidual}^t_{i - 2}, \hat{\InterResidual}^t_{i - 3} \right] \right) \right) \\
	\forall \ i \geq 3 \land t \bmod 2 = 1.
\end{multline}
In this case, the reference images are residual signals as well.
Again, as soon as this residual is decoded again, it serves as reference residual for the next three channels.

\subsubsection{Evaluation}

Our novel hyperspectral video database can directly be used to evaluate the gains of the proposed hyperspectral video coder.
For the evaluation, only the videos of the center camera are used.
The Quantization Parameters (QPs) for evaluation are set to 22, 27, 32 and 37.
These QPs are extracted from the common test conditions for HEVC and thus common practice\cite{hevc_ctc_2018}.

In general, different coders are evaluated by setting up rate-distortion curves, i.e., a plot that shows the data rate in terms of bitrate on the x-axis and distortion, here in terms of PSNR, on the y-axis.
The different aforementioned QPs are datapoints for this curve.
Between these data points the curve is interpolated.
Thus, if the curve of a second coder is to the top left compared to the curve of the base coder, it achieves a better PSNR for the same bitrate, or vice versa, a lower bit rate for the same PSNR.
This average distance between rate-distortion curves can be measured using the Bjontegaard-Delta (BD) for the rate direction and PSNR direction\cite{bjontegaard_calculation_2001}.
The results for each scene are summarized in Table\ \ref{tab:eval_coding}, which shows the BD rate and PSNR.
The proposed hyperspectral video coder is first compared to the all-intra coder PRIBP.
For six out of seven scenes, the proposed coder outperforms the intra procedure.
PWC-Net has problems estimating the motion properly for the scene \textit{forest}, since there are a lot of leaves, which leads to a high frequency change between background and leaves.
Thus, the motion estimation would have to output a motion field with a very high spatial resolution to enhance the coder.
In this case, PWC-Net does not achieve this and even has a bad influence on the coding result.
On average, the novel hyperspectral coder saves roughly 3\% in comparison to PRIBP. Note that this average is calculated over all sizes and PSNRs and not over the rate and PSNR values from the table.
Thus, videos with a large file size influences this average more.

Moreover, the proposed coding technique is also compared to the random access mode of High Efficiency Video Coding (HEVC), thus an RGB inter coder.
For that, every three consecutive channels were merged into an RGB video.
The remaining channels are coded as grayscale videos.
For 31 hyperspectral channels, this results in 10 RGB videos and 1 grayscale video to code for the standard HEVC.
The results vary much more in comparison, however, the average bit rate saving of the proposed hyperspectral video coder is over 50\% in comparison the inter mode of HEVC.
It is noticeable that HEVC has problems with high frequency content in comparison, since the scenes \textit{family house}, \textit{outdoor} and \textit{forest} are compressed much worse.
On the other side, for scenes with mostly homogeneous regions and movements, HEVC even outperforms the proposed hyperspectral video coder.

\begin{table}[t]
	\centering
	\renewcommand{\arraystretch}{1.5}
	\caption{Evaluation of our novel inter hyperspectral coding against all-intra hyperspectral coding PRIBP and HEVC inter coding (random access). The values are given in Bjontegaard-Delta rate and PSNR, respectively. Lower is better for rate, while higher is better for PSNR.}
	\label{tab:eval_coding}
	\begin{tabular}{c|c|c|c|c}
		& \multicolumn{2}{c|}{PRIBP} & \multicolumn{2}{c}{HEVC Inter Coding} \\
		& BD Rate   & BD PSNR   & BD Rate   & BD PSNR   \\ \hline
		Fam. house          & -10.41\%  & 0.59 dB   & -54.55\%  & 4.02 dB   \\ \hline
		M. seaport          & -7.15\%   & 0.35 dB   & -22.82\%  & 1.16 dB   \\ \hline
		City                & -9.29\%   & 0.44 dB   & -19.36\%  & 0.92 dB   \\ \hline
		Outdoor             & -9.22\%   & 0.47 dB   & -65.81\%  & 4.82 dB   \\ \hline
		Forest              & 13.24\%   & -0.59 dB  & -73.49\%  & 6.09 dB   \\ \hline
		Indoor              & -6.77\%   & 0.22 dB   & 12.71\%   & -0.38 dB  \\ \hline
		Lab                 & -9.21\%   & 0.57 dB   & 42.61\%   & -1.85 dB  \\ \hline \hline
		Average             & -3.36\%   & 0.15 dB   & -52.46\%  & 3.27 dB
	\end{tabular}
	\vspace*{-0.3cm}
\end{table}

\subsection{Other Applications}
Other applications include spectral reconstruction\cite{sippel_structure-preserving_2020}.
Spectral reconstruction has the goal of reconstructing light spectra from multispectral images or even RGB images.
The database can also be used for cross-spectral depth estimation\cite{genser_deep_2020}, where the disparity between different views, which record different ranges in the spectrum, is estimated, e.g., using the multispectral camera array shown in \fig\ref{fig:camsi}.
Furthermore, one can also evaluate different camera setups like trinocular arrangements\cite{mozerov_trinocular_2009}.
Moreover, depth estimation algorithms and image enhancement procedures that rely on grayscale and RGB cameras\cite{jeon_stereo_2016} can be used.
In a similar direction steer studies that investigate proper color to grayscale conversions for these types of algorithms\cite{benedetti_color_2012}.
A related topic to stereo matching is scene reconstruction using different cameras\cite{vedaldi_atlas_2020}.

In general, different fusion techniques related to depth sensors can be evaluated.
A very obvious example is depth estimation aided by a depth map sensor, e.g., a LIDAR sensor\cite{gao_object_2018}.
Furthermore, a depth sensor can be used to help upsampling\cite{park_high-quality_2014} as well.
An even better fit to this database is the fusion of RGB images, hyperspectral recordings and LIDAR data for urban land cover classification\cite{hansch_fusion_2021}.

Of course reconstruction algorithms for different types of hyperspectral cameras can be easily evaluated like demosaicing for multispectral filter arrays\cite{feng_mosaic_2021}.
Combinations of these techniques also can be evaluated, e.g., a mixture of multispectral filter arrays and camera arrays.

Finally, multispectral and hyperspectral denoising algorithms can be evaluated, e.g., techniques based on sparse matrix decomposition\cite{xie_hyperspectral_2020}.
The noise can be even further reduced exploiting the temporal correlation between frames, which can be evaluated using the proposed database.

\section{Conclusion}
\label{sec:conclusion}
This paper introduced a novel synthetic hyperspectral video database containing seven scenes.
All scenes were rendered using a camera array with nine cameras arranged in a three times three grid.
This camera array moves through the scenes for 30 frames.
Since depth maps are provided for every scene, camera and frame, this database provides a high resolution in all three spatial as well as spectral and temporal dimension of the corresponding scene.
Therefore, this database can serve as validation database for many applications ranging from spectral reconstruction over diverse stereo matching problems like cross spectral disparity estimation to sensor fusion problem like combining hyperspectral data, RGB images and data from depth sensing devices.
Finally, two applications were covered in detail by exploiting the temporal dimension of the data.
The proposed two novel algorithms outperformed its non-temporal versions.
Furthermore, it proved that this new database can be used to evaluate algorithms in diverse image processing tasks.

\begin{backmatter}
 \bmsection{Funding} The authors gratefully acknowledge that this work has been supported by the Deutsche Forschungsgemeinschaft (DFG, German Research Foundation) under project number 491814627.
 \bmsection{Disclosures} The authors declare no conflicts of interest.
 \bmsection{Data availability} Data underlying the results presented in this paper are available in Dataset 1, Ref. \cite{hyvid}.
\end{backmatter}

\bibliography{refs}

\begin{thebibliography}{10}
\newcommand{\enquote}[1]{``#1''}

\bibitem{hagen_review_2013}
N.~A. Hagen and M.~W. Kudenov, \enquote{{Review of snapshot spectral imaging
  technologies},} {\protect\JournalTitle{Optical Engineering}} \textbf{52}, 1
  -- 23 (2013).

\bibitem{garaba_airborne_2018}
S.~P. Garaba and H.~M. Dierssen, \enquote{{An airborne remote sensing case
  study of synthetic hydrocarbon detection using short wave infrared absorption
  features identified from marine-harvested macro- and microplastics},}
  {\protect\JournalTitle{Remote Sensing of Environment}} \textbf{205}, 224--235
  (2018).

\bibitem{williams_classification_2016}
P.~J. Williams and S.~Kucheryavskiy, \enquote{{Classification of maize kernels
  using NIR hyperspectral imaging},} {\protect\JournalTitle{Food Chemistry}}
  \textbf{209}, 131--138 (2016).

\bibitem{edelman_hyperspectral_2012}
G.~Edelman, E.~Gaston, T.~{van Leeuwen}, P.~Cullen, and M.~Aalders,
  \enquote{{Hyperspectral imaging for non-contact analysis of forensic
  traces},} {\protect\JournalTitle{Forensic Science International}}
  \textbf{223}, 28--39 (2012).

\bibitem{xiong_material_2020}
F.~Xiong, J.~Zhou, and Y.~Qian, \enquote{{Material Based Object Tracking in
  Hyperspectral Videos},} {\protect\JournalTitle{IEEE Transactions on Image
  Processing}} \textbf{29}, 3719--3733 (2020).

\bibitem{han_vivo_2016}
Z.~Han, A.~Zhang, X.~Wang, Z.~Sun, M.~D. Wang, and T.~Xie, \enquote{{In vivo
  use of hyperspectral imaging to develop a noncontact endoscopic diagnosis
  support system for malignant colorectal tumors},}
  {\protect\JournalTitle{Journal of Biomedical Optics}} \textbf{21}, 016001
  (2016).

\bibitem{yasuma_generalized_2010}
F.~Yasuma, T.~Mitsunaga, D.~Iso, and S.~K. Nayar, \enquote{{Generalized
  Assorted Pixel Camera: Postcapture Control of Resolution, Dynamic Range, and
  Spectrum},} {\protect\JournalTitle{IEEE Transactions on Image Processing}}
  \textbf{19}, 2241--2253 (2010).

\bibitem{eckhard_outdoor_2015}
J.~Eckhard, T.~Eckhard, E.~M. Valero, J.~L. Nieves, and E.~G. Contreras,
  \enquote{{Outdoor scene reflectance measurements using a Bragg-grating-based
  hyperspectral imager},} {\protect\JournalTitle{Appl. Opt.}} \textbf{54},
  D15--D24 (2015).

\bibitem{hordley_multi-spectral_2004}
S.~Hordley, G.~Finalyson, and P.~Morovic, \enquote{{A multi-spectral image
  database and its application to image rendering across illumination},} in
  \emph{Third International Conference on Image and Graphics (ICIG'04),}
  (2004), pp. 394--397.

\bibitem{larabi_database_2015}
S.~L. Moan, S.~T. George, M.~Pedersen, J.~Blahová, and J.~Y. Hardeberg,
  \enquote{{A database for spectral image quality},} in \emph{Image Quality and
  System Performance XII,}  vol. 9396 M.-C. Larabi and S.~Triantaphillidou,
  eds., International Society for Optics and Photonics (SPIE, 2015), pp. 225 --
  232.

\bibitem{foster_frequency_2006}
D.~H. Foster, K.~Amano, S.~M.~C. Nascimento, and M.~J. Foster,
  \enquote{{Frequency of metamerism in natural scenes},}
  {\protect\JournalTitle{J. Opt. Soc. Am. A}} \textbf{23}, 2359--2372 (2006).

\bibitem{chakrabarti_statistics_2011}
T.~Zickler and A.~Chakrabarti, \enquote{{Statistics of real-world hyperspectral
  images},} in \emph{2013 IEEE Conference on Computer Vision and Pattern
  Recognition,}  (IEEE Computer Society, Los Alamitos, CA, USA, 2011), pp.
  193--200.

\bibitem{leibe_sparse_2016}
B.~Arad and O.~Ben-Shahar, \enquote{{Sparse Recovery of Hyperspectral Signal
  from Natural RGB Images},} in \emph{Computer Vision -- ECCV 2016,}  B.~Leibe,
  J.~Matas, N.~Sebe, and M.~Welling, eds. (Springer International Publishing,
  Cham, 2016), pp. 19--34.

\bibitem{mian_hyperspectral_2012}
A.~Mian and R.~Hartley, \enquote{Hyperspectral video restoration using optical
  flow and sparse coding,} {\protect\JournalTitle{Opt. Express}} \textbf{20},
  10658--10673 (2012).

\bibitem{gomez-chova_correction_2008}
L.~G\'{o}mez-Chova, L.~Alonso, L.~Guanter, G.~Camps-Valls, J.~Calpe, and
  J.~Moreno, \enquote{{Correction of systematic spatial noise in push-broom
  hyperspectral sensors: application to CHRIS/PROBA images},}
  {\protect\JournalTitle{Appl. Opt.}} \textbf{47}, F46--F60 (2008).

\bibitem{koenig_practice_1998}
F.~Koenig and W.~Praefcke, \enquote{{Practice of multispectral image
  acquisition},} in \emph{Electronic Imaging: Processing, Printing, and
  Publishing in Color,}  vol. 3409 J.~Bares, ed., International Society for
  Optics and Photonics (SPIE, 1998), pp. 34 -- 41.

\bibitem{gat_development_2006}
N.~Gat, G.~Scriven, J.~Garman, M.~D. Li, and J.~Zhang, \enquote{{Development of
  four-dimensional imaging spectrometers (4D-IS)},} in \emph{Imaging
  Spectrometry XI,}  vol. 6302 S.~S. Shen and P.~E. Lewis, eds., International
  Society for Optics and Photonics (SPIE, 2006), pp. 179 -- 189.

\bibitem{descour_demonstration_1997}
M.~R. Descour, C.~E. Volin, E.~L. Dereniak, T.~M. Gleeson, M.~F. Hopkins, D.~W.
  Wilson, and P.~D. Maker, \enquote{{Demonstration of a computed-tomography
  imaging spectrometer using acomputer-generated hologram disperser},}
  {\protect\JournalTitle{Appl. Opt.}} \textbf{36}, 3694--3698 (1997).

\bibitem{matchett_volume_2007}
J.~D. Matchett, R.~I. Billmers, E.~J. Billmers, and M.~E. Ludwig,
  \enquote{{Volume holographic beam splitter for hyperspectral imaging
  applications},} in \emph{Novel Optical Systems Design and Optimization X,}
  vol. 6668 R.~J. Koshel and G.~G. Gregory, eds., International Society for
  Optics and Photonics (SPIE, 2007), pp. 191 -- 198.

\bibitem{monno_practical_2015}
Y.~Monno, S.~Kikuchi, M.~Tanaka, and M.~Okutomi, \enquote{{A Practical One-Shot
  Multispectral Imaging System Using a Single Image Sensor},}
  {\protect\JournalTitle{IEEE Transactions on Image Processing}} \textbf{24},
  3048--3059 (2015).

\bibitem{shogenji_multispectral_2004}
R.~Shogenji, Y.~Kitamura, K.~Yamada, S.~Miyatake, and J.~Tanida,
  \enquote{{Multispectral imaging using compact compound optics},}
  {\protect\JournalTitle{Opt. Express}} \textbf{12}, 1643--1655 (2004).

\bibitem{gehm_single-shot_2007}
M.~E. Gehm, R.~John, D.~J. Brady, R.~M. Willett, and T.~J. Schulz,
  \enquote{{Single-shot compressive spectral imaging with a dual-disperser
  architecture},} {\protect\JournalTitle{Opt. Express}} \textbf{15},
  14013--14027 (2007).

\bibitem{genser_camera_2020}
N.~Genser, J.~Seiler, and A.~Kaup, \enquote{{Camera Array for Multi-Spectral
  Imaging},} {\protect\JournalTitle{IEEE Transactions on Image Processing}}
  \textbf{29}, 9234--9249 (2020).

\bibitem{blender}
B.~O. Community, \emph{{Blender - a 3D modelling and rendering package}},
  Blender Foundation, Stichting Blender Foundation, Amsterdam (2018).

\bibitem{purcell_ray_2005}
T.~J. Purcell, I.~Buck, W.~R. Mark, and P.~Hanrahan, \enquote{{Ray Tracing on
  Programmable Graphics Hardware},} {\protect\JournalTitle{ACM Trans. Graph.}}
  \textbf{21}, 703–712 (2002).

\bibitem{buades_non-local_2011}
A.~Buades, B.~Coll, and J.-M. Morel, \enquote{{Non-Local Means Denoising},}
  {\protect\JournalTitle{{Image Processing On Line}}} \textbf{1}, 208--212
  (2011).

\bibitem{mccann_veiling_2007}
J.~J. McCann and A.~Rizzi, \enquote{{Veiling glare: the dynamic range limit of
  HDR images},} in \emph{Human Vision and Electronic Imaging XII,}  vol. 6492
  B.~E. Rogowitz, T.~N. Pappas, and S.~J. Daly, eds., International Society for
  Optics and Photonics (SPIE, 2007), p. 649213.

\bibitem{rosenhauer_image_1968}
K.~Rosenhauer and K.~Rosenbruch, \enquote{Flare and optical transfer function,}
  {\protect\JournalTitle{Appl. Opt.}} \textbf{7}, 283--287 (1968).

\bibitem{waller_phase_2010}
L.~Waller, S.~S. Kou, C.~J.~R. Sheppard, and G.~Barbastathis, \enquote{Phase
  from chromatic aberrations,} {\protect\JournalTitle{Opt. Express}}
  \textbf{18}, 22817--22825 (2010).

\bibitem{hytti_noise_2006}
H.~T. Hytti, \enquote{{Characterization of digital image noise properties based
  on RAW data},} in \emph{Image Quality and System Performance III,}  vol. 6059
  L.~C. Cui and Y.~Miyake, eds., International Society for Optics and Photonics
  (SPIE, 2006), p. 60590A.

\bibitem{sippel_spatio-spectral_2021}
F.~Sippel, J.~Seiler, and A.~Kaup, \enquote{{Spatio-spectral Image
  Reconstruction Using Non-local Filtering},} in \emph{2021 International
  Conference on Visual Communications and Image Processing (VCIP),}  (2021),
  pp. 1--5.

\bibitem{sippel_structure-preserving_2020}
F.~Sippel, J.~Seiler, N.~Genser, and A.~Kaup, \enquote{{Structure-preserving
  spectral reflectance estimation using guided filtering},}
  {\protect\JournalTitle{J. Opt. Soc. Am. A}} \textbf{37}, 1695--1710 (2020).

\bibitem{meyer_multispectral_2020}
A.~Meyer, N.~Genser, and A.~Kaup, \enquote{{Multispectral Image Compression
  Based on HEVC Using Pel-Recursive Inter-Band Prediction},} in \emph{2020 IEEE
  22nd International Workshop on Multimedia Signal Processing (MMSP),}  (2020),
  pp. 1--6.

\bibitem{sullivan_overview_2012}
G.~J. Sullivan, J.-R. Ohm, W.-J. Han, and T.~Wiegand, \enquote{{Overview of the
  High Efficiency Video Coding (HEVC) Standard},} {\protect\JournalTitle{IEEE
  Transactions on Circuits and Systems for Video Technology}} \textbf{22},
  1649--1668 (2012).

\bibitem{sun_pwc-net_2018}
D.~Sun, X.~Yang, M.-Y. Liu, and J.~Kautz, \enquote{{PWC-Net: CNNs for Optical
  Flow Using Pyramid, Warping, and Cost Volume},} in \emph{2018 IEEE/CVF
  Conference on Computer Vision and Pattern Recognition,}  (2018), pp.
  8934--8943.

\bibitem{hevc_ctc_2018}
J.~Boyce, K.~Suehring, X.~Li, and V.~Seregin, \enquote{{JVET-J1010: JVET common
  test conditions and software reference configurations},}  (2018).

\bibitem{bjontegaard_calculation_2001}
G.~Bjøntegaard, \enquote{{Calculation of average PSNR differences between RD
  curves},} in \emph{{VCEG-M33, document},}  (2001).

\bibitem{genser_deep_2020}
N.~Genser, A.~Spruck, J.~Seiler, and A.~Kaup, \enquote{{Deep Learning Based
  Cross-Spectral Disparity Estimation For Stereo Imaging},} in \emph{2020 IEEE
  International Conference on Image Processing (ICIP),}  (2020), pp.
  2536--2540.

\bibitem{mozerov_trinocular_2009}
M.~Mozerov, J.~Gonzàlez, X.~Roca, and J.~J. Villanueva, \enquote{{Trinocular
  stereo matching with composite disparity space image},} in \emph{2009 16th
  IEEE International Conference on Image Processing (ICIP),}  (2009), pp.
  2089--2092.

\bibitem{jeon_stereo_2016}
H.-G. Jeon, J.-Y. Lee, S.~Im, H.~Ha, and I.~S. Kweon, \enquote{{Stereo Matching
  with Color and Monochrome Cameras in Low-Light Conditions},} in \emph{2016
  IEEE Conference on Computer Vision and Pattern Recognition (CVPR),}  (2016),
  pp. 4086--4094.

\bibitem{benedetti_color_2012}
L.~Benedetti, M.~Corsini, P.~Cignoni, M.~Callieri, and R.~Scopigno,
  \enquote{Color to gray conversions in the context of stereo matching
  algorithms: {An} analysis and comparison of current methods and an ad-hoc
  theoretically-motivated technique for image matching,}
  {\protect\JournalTitle{Machine Vision and Applications}} \textbf{23},
  327--348 (2012).

\bibitem{vedaldi_atlas_2020}
Z.~Murez, T.~van As, J.~Bartolozzi, A.~Sinha, V.~Badrinarayanan, and
  A.~Rabinovich, \enquote{{Atlas: End-to-End 3D Scene Reconstruction from Posed
  Images},} in \emph{Computer Vision -- ECCV 2020,}  A.~Vedaldi, H.~Bischof,
  T.~Brox, and J.-M. Frahm, eds. (Springer International Publishing, Cham,
  2020), pp. 414--431.

\bibitem{gao_object_2018}
H.~Gao, B.~Cheng, J.~Wang, K.~Li, J.~Zhao, and D.~Li, \enquote{{Object
  Classification Using CNN-Based Fusion of Vision and LIDAR in Autonomous
  Vehicle Environment},} {\protect\JournalTitle{IEEE Transactions on Industrial
  Informatics}} \textbf{14}, 4224--4231 (2018).

\bibitem{park_high-quality_2014}
J.~Park, H.~Kim, Y.-W. Tai, M.~S. Brown, and I.~S. Kweon,
  \enquote{{High-Quality Depth Map Upsampling and Completion for RGB-D
  Cameras},} {\protect\JournalTitle{IEEE Transactions on Image Processing}}
  \textbf{23}, 5559--5572 (2014).

\bibitem{hansch_fusion_2021}
R.~Hänsch and O.~Hellwich, \enquote{{Fusion of Multispectral LiDAR,
  Hyperspectral, and RGB Data for Urban Land Cover Classification},}
  {\protect\JournalTitle{IEEE Geoscience and Remote Sensing Letters}}
  \textbf{18}, 366--370 (2021).

\bibitem{feng_mosaic_2021}
K.~Feng, Y.~Zhao, J.~C.-W. Chan, S.~G. Kong, X.~Zhang, and B.~Wang,
  \enquote{{Mosaic Convolution-Attention Network for Demosaicing Multispectral
  Filter Array Images},} {\protect\JournalTitle{IEEE Transactions on
  Computational Imaging}} \textbf{7}, 864--878 (2021).

\bibitem{xie_hyperspectral_2020}
T.~Xie, S.~Li, and B.~Sun, \enquote{{Hyperspectral Images Denoising via
  Nonconvex Regularized Low-Rank and Sparse Matrix Decomposition},}
  {\protect\JournalTitle{IEEE Transactions on Image Processing}} \textbf{29},
  44--56 (2020).

\bibitem{hyvid}
\enquote{{HyViD},} \url{https://github.com/FAU-LMS/HyViD}.

\end{thebibliography}

\end{document}